\documentclass[aps,pra,showpacs, twocolumn, amsfonts, amsmath]{revtex4}

\usepackage{amsmath}
\usepackage{amssymb}
\usepackage{epsf}
\usepackage{epsfig}
\usepackage{graphicx}
\usepackage[english]{babel}
\usepackage[latin1]{inputenc}
\usepackage{rotating}
\usepackage{color}
\usepackage{slashed}
\usepackage{txfonts}

\newcommand{\bea}{\begin{eqnarray}}
\newcommand{\eea}{\end{eqnarray}}

\begin{document}

\title{Real-time effective-action approach to the Anderson quantum dot}

\author{D\'enes Sexty}
\author{Thomas Gasenzer}
\author{Jan Pawlowski} 

\affiliation{Institut f\"ur Theoretische Physik,
             Ruprecht-Karls-Universit\"at Heidelberg,
             Philosophenweg~16,
             69120~Heidelberg, Germany}
\affiliation{ExtreMe Matter Institute EMMI,
             GSI Helmholtzzentrum f\"ur Schwerionenforschung GmbH, 
             Planckstra\ss e~1, 
             64291~Darmstadt, Germany}

\date{\today}

\begin{abstract}

  The non-equilibrium time evolution of an Anderson quantum dot is
  investigated. The quantum dot is coupled between two leads forming a
  chemical-potential gradient. We use
  Kadanoff-Baym dynamic equations within a non-perturbative
  resummation of the s-channel bubble chains.  The effect of the
  resummation leads to the introduction of a frequency-dependent
  4-point vertex. The tunneling to the leads is taken into account
  exactly.  The method allows the determination of the transient as
  well as stationary transport through the quantum dot, and results
  are compared with different schemes discussed in the literature
  (fRG, ISPI, tDMRG and QMC).

\end{abstract}


\maketitle

\section{Introduction}\label{Introduction}
Due to the recent advances in nanoscale technology, it has become
possible to study electron transport experimentally in nano-devices
such as single molecules, artificially designed quantum dots or
nanotubes
\cite{molelecbook,reichert2002,Smit2002,Park2000,Galperin2007}.  This
naturally incited new interest, and quantum dot models have
been the subject of much theoretical effort recently. The quantum dot also
serves as a good playground for theoretical methods: it allows for
many-body effects as well as 
non-equilibrium phenomena in a regime where linear response theory no longer applies.
The description of co-tunneling processes such as the Kondo effect and 
the emergence of the Kondo scale have proven to be especially
challenging. This is the case, in particular, for the exponential coupling dependence of the Kondo scale.

The Kondo regime has been investigated perturbatively using
Fermi liquid theory \cite{Oguri2001}, exploiting integrability 
of the Anderson model \cite{Konik2002}, or the non-crossing approximation
at infinite local coupling $U$ \cite{Wingreen1994}. 
Several renormalization-group methods have been used to investigate the
stationary state of the system: the perturbative real-time
renormalization group (RG) \cite{Schoeller2000}, non-equilibrium 
extensions of perturbative RG \cite{Kaminski2000,Rosch2001,Rosch2003},
and the functional RG
approach in its generalization to non-equilibrium situations
\cite{Andergassen2006,Gezzi2007,Jakobs2007,Gasenzer:2008zz,Gasenzer2008a,Jakobs2010,Karrasch2010}.
Bethe-Salpeter equations in so-called parquet approximation
\cite{Janis1999} were analysed with respect to the Kondo resonance in Refs. 
\cite{Janis2007,Janis2008,Aug2010}.
The numerical renormalization group (NRG) method has also 
been successful in describing such impurity systems, and it has 
been generalized to describe time-dependent non-equilibrium systems 
\cite{Anders2005,Anders2006,Roosen2008,Schmitt2010}.

Another line of thought is represented by the quantum Monte 
Carlo method (QMC)
\cite{egger2000}, which is
 numerically exact, but only short simulation times are possible 
since the method is plagued by the sign problem at small temperatures.

The transient behaviour of the system as well as the stationary
state have also been explored by methods such as the time dependent
density matrix renormalisation group (tDMRG)
\cite{Daley2004,White2004,HM2009,Boulat2008} which
allows the simulation of the time evolution of pure states, and the
iterative real-time summation of the path integral (ISPI) \cite{ispi} which
is numerically exact, but depends on the correlation time of the system
being small.
Several of these theoretical methods were recently compared in 
\cite{eckel2010}, which also contains a concise list of related previous  
studies.

In this article we use the two-particle-irreducible (2PI) effective action 
or $\Phi$-functional
\cite{luttingerward,Baym1962,Cornwall1974a}, to derive Kadanoff-Baym equations
of motion describing the transient and stationary transport through 
a quantum dot.
The power of this method is that 
symmetries of the system are 
conserved during the time evolution, i.e., particle 
number and energy conservation is satisfied.
This method has been successfully applied to the thermalisation of relativistic  and nonrelativistic systems
of bosonic and fermionic gases
\cite{Berges:2001fi,Berges2003a,Gasenzer:2005ze,Arrizabalaga:2005tf,Berges:2007ym,Kronenwett:2010dx,Giraud:2009tn}.

We apply the 2PI method to the single-impurity 
Anderson model of a quantum dot within two different truncations: 
the loop expansion up to
two-loop order (in the self-energy), and a resummation, which
includes the contribution of the spin-aligned bubble chains summed to all
orders. As shown below, the effect of this s-channel resummation can be seen 
as using a frequency dependent 4-point vertex.
  The leads can be taken into account exactly in this 
formalism: integrating them out they give a contribution to 
the self-energy of the dot-electrons. 

After benchmarking our method against exact solutions for an isolated quantum dot we compare our results for the transient and stationary behavior of the dot coupled to the leads with results found in other schemes, including fRG, ISPI, tDMRG, and QMC calculations.
While finding good agreement in the long-time stationary limit with the behavior found with these schemes our approach also allows to treat the transient build-up of the current through the dot.
Comparisons of the stationary current for different interactions show good agreement with the alternative schemes over a large range of interaction strengths.
These results underline the necessity of a nonperturbative resummation.

The structure of the paper is as follows: The Anderson model 
and its ingredients are defined in Section \ref{modelsect}. In
Section \ref{2pisect} we summarize the formulation of the 2PI effective-action approach 
for nonrelativistic Fermions on an isolated quantum dot.
We introduce the formulation for the dot coupled to leads in Section \ref{leadsect}.  
In Section \ref{exasect} we compare our method with exact results, for cases
where an analytical treatment is possible.  In Section \ref{ressect} we
present our results for the build-up of the current and for its stationary characteristics, 
and compare these to results in the literature. 
We draw our conclusions in Section \ref{consect}.

\begin{figure}[t]
\begin{center} 
\includegraphics[width=0.45 \textwidth]{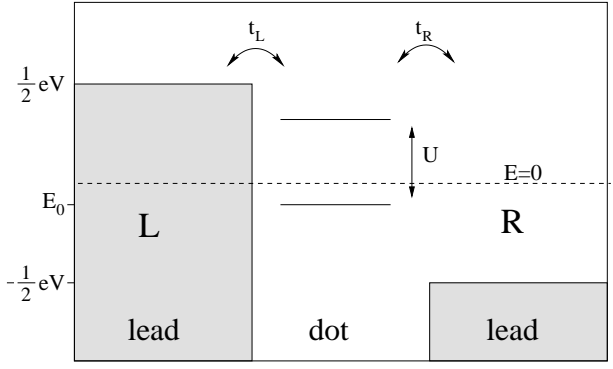}
\caption{ The schematic representation of the energy levels in the 
Anderson dot system, for the zero magnetic field case. The lower line of the 
dot represents the energy level of the first electron 
occupying the dot, the upper line is the level of the second electron
occupying the dot.
\label{fig:dot}}
\end{center}
\end{figure}

\section{The Anderson model }
\label{modelsect}

The Anderson Hamiltonian describing the electrons on the quantum dot and on 
the leads reads

\bea \label{andersonh}
 H 
 &=& H_\textrm{dot} + H_\textrm{leads}+ H_\textrm{tunnel} 
 = \sum_\sigma E_{0\sigma} n_\sigma + U n_{\uparrow} n_{\downarrow} + 
 \nonumber\\
&&+\ \sum_{k p \sigma }  \epsilon_{kp}  c^\dagger_{kp\sigma}  c_{kp\sigma} 
- \sum_{kp\sigma} (t_p c^\dagger_{kp\sigma} d_\sigma + t_p^* d^\dagger_\sigma c_{k p \sigma} )
\eea
where $\sigma= \pm 1/2 \simeq \ \uparrow , \downarrow$ is the spin index, 
$p=\pm \simeq L/R $ labels the leads on the left and right, respectively, 
and $k$ is the
index of the spectrum of the lead electrons.  The leads on the left
and right have a chemical potential $ \mu_p= p eV/2$ which models the
bias voltage. For the determination of the energies 
$\epsilon_{kp}$, 
see Sect.~\ref{leadsect}.
  The occupation-number operator on the dot is written
as: $ n_\sigma= d^\dagger_\sigma d_\sigma $.  The dot electrons have
an interaction term $\sim U$ due to Coulomb repulsion.
The one-body energy is
$E_{0\sigma} = E_0 + \sigma B $. The first term is controlled
experimentally through the gate voltage whereas the second (Zeeman) term 
by the magnetic field $B$. The energy levels of the system are 
represented in Fig.~\ref{fig:dot}.

The tunneling between the dot and the leads is quantified by the tunneling
strength $t_p$. 
 We will assume that the leads are in thermal equilibrium
 at some temperature $T$ and use the usual wide-band limit with a 
constant density of states $\rho$ around the Fermi surface.
In this work we use a symmetric 
tunneling rate $ \tau =t_L=t_R $. The dimensionful quantities of the system 
can be expressed in units of the hybridization $ \Gamma= 2 \pi |\tau|^2 \rho $ 
which quantifies the dressing of the dot by the leads.

We will use, in the following, a 
Lagrangian or action formulation of the Anderson quantum field model.
The contributions to the action corresponding to those forming 
the Hamiltonian (\ref{andersonh}) are given by
\bea 
S_\textrm{dot}
&=& \int_C dt \sum_\sigma d^\dagger_\sigma ( i \partial_t -E_{0\sigma}) d_\sigma
 - U d^\dagger_{\uparrow} d_{\uparrow} d^\dagger_{\downarrow} d_{\downarrow},  
 \\ \label{S:leads}
  S_\textrm{leads}
  &=& \int _C dt  \sum_{kp\sigma} c^\dagger_{kp\sigma} ( i \partial_t - 
 \epsilon_{kp} ) c_{kp\sigma}, 
 \\ \label{S:tunnel}
 S_\textrm{tunnel}
 &=& \int _C dt \sum_{kp\sigma} ( t_p c^\dagger_{kp\sigma} d_\sigma + 
 t^*_p d^\dagger_\sigma c_{kp\sigma} ).
\eea

Our most important observable is the time dependent current through the dot
\bea
I(t) = - {i e\over 2 } \sum_{kp\sigma} \left(
p t_p \langle c^\dagger_{kp\sigma} d_\sigma \rangle 
 - p t^*_p \langle d^\dagger_\sigma c_{kp\sigma} \rangle \right).
\eea 
This can also be written as $I=(I_L-I_R)/2$, where $I_p=-e \dot N_p(t)$ and 
$N_p(t)= \langle \sum_{k\sigma} c^\dagger_{pk\sigma} c_{pk\sigma} \rangle $ the 
number of electrons on the leads. The stationary current can simply be 
obtained by waiting for the transient behavior to die out, such that
the system is sufficiently close to the final, stationary state.

An important scale of the physical processes described by the Anderson model is given by
the Kondo temperature. 
This temperature marks the onset of the Kondo effect which is due to the formation 
of singlet states of intinerant and localised dot fermions and characterised by a rising resistivity
of the dot at low temperatures. The Kondo temperature is found by simple renormalisation-group
arguments to be
\bea \label{kondotemp}
T_K=\sqrt{ U \Gamma \over 2 } \exp \left(  - \pi U \over 8 \Gamma \right).
\eea
 for the particle-hole symmetric system, where $E_0=-U/2$.

\section{2PI formalism for the isolated dot}
\label{2pisect}

In this paper, we study Kadanoff-Baym-type dynamic equations for correlation functions describing transport 
through the Anderson quantum dot.
We derive these equations using the two-particle irreducible (2PI) effective-action 
formalism, also known as $\Phi$-derivable approach.
In the following, we summarize known basics about the 2PI effective-action approach 
\cite{luttingerward,Baym1962,Cornwall1974a} and introduce 
specific details of its implementation for the Anderson model.

Since we will be interested in initial value problems, specifically, in the evolution of multi-time correlation functions
starting from values given by some initial state, it is convenient to work in the Heisenberg picture.
In the corresponding functional-integral formulation, the time integrations in the action are defined 
to run along a Schwinger-Keldysh contour $\cal C$, which goes from the initial time $t_0$ to
some final time $t$, and then back to $t_0$ \cite{Keldysh1964a,Chou:1984es}. 

The Kadanoff-Baym equations will yield the evolution of the time-ordered two-point function:
\begin{equation}
\label{eq:DefD}
 D_{\sigma\lambda}(t,t') = \Theta_C (t-t') \langle d_\sigma(t) d_\lambda^\dagger (t') \rangle
       - \Theta_C (t'-t)  \langle d_\lambda^\dagger(t') d_\sigma(t) \rangle\hspace{0.2cm}
\end{equation}
where $\Theta_C (t-t')$ is a $\Theta$-function on the
Schwinger-Keldysh contour, and evaluates to 1 (0) if 
$t$ is later (earlier) to $t'$.

\begin{figure}[t]
\begin{center} 
\includegraphics[width=0.45 \textwidth]{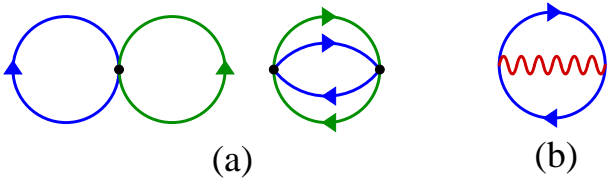}
\caption{(color online) 2PI diagrams of the loop expansion of $\Gamma_2 [D]$.
(a) The two lowest-order diagrams of the loop
expansion described in sections 
\ref{meansec} and \ref{basketsec}, with each vertex coupling up spins
to down spins. Black dots represent the bare vertex $\sim U$, 
blue (green) solid lines the up (down) spin propagator $D_{\sigma\sigma} , \ 
\sigma= \uparrow, \downarrow $.
(b) Diagram representing the resummation approximation explained in 
Sect. \ref{resumsec}. The 
wiggly line is the scalar propagator $G_{11}$. 
This propagator is represented as a sum of bubble-chain diagrams, see (\ref{eq:scalarProp}).
The same diagram with 
down spins also appears in $\Gamma_2[D] $.
\label{fig:loops}}
\end{center}
\end{figure}

We begin with the simpler case of the isolated quantum dot.
Without leads the 2PI effective action \cite{Cornwall1974a} reads
\bea 
 \label{eq:2PIEAwoleads}
 \Gamma[ D] = - i \textrm{Tr} 
 \left[ \textrm{ln} D^{-1} + D_0^{-1} D \right] + \Gamma_2 [D] + \textrm{const.}
\eea 
Here $\Gamma_2[D]$ can be written as the sum of all closed 2PI diagrams constructed from bare 
vertices and full propagators.
Two-particle irreducible are those diagrams which do not fall apart
upon cutting two lines \cite{Cornwall1974a}.
As an example we show, in Fig~\ref{fig:loops}a, the two diagrams of 
lowest order in the bare coupling $U$, represented as a black dot.
Full lines denote different spin components of the full propagator or two-point function $D$.
In (\ref{eq:2PIEAwoleads}), the free inverse propagator is given by
\bea i D_{0,\sigma\lambda}^{-1} (t,t') = ( i \partial_t - E_{0\sigma}) \delta_C (t-t') \delta_{\sigma\lambda}.
\eea
The Kadanoff-Baym equations of motion are calculated from the stationarity
conditions for the action
\bea { \delta \Gamma [D ] \over \delta D_{\lambda \sigma }(t' ,t )} = 0 .
\eea
This equation gives the well-known Dyson equation $D^{-1}_{\sigma\lambda}(t,t')=
D_{0,\sigma\lambda}^{-1}(t,t') - \Sigma_{\sigma\lambda}(t,t') $, where the self-energy is given by
\bea \Sigma_{\sigma\lambda}(t,t') = -i { \delta \Gamma_2 [D] \over
  \delta D_{\lambda\sigma} (t',t) }. \eea
We use the decomposition of the time-ordered two-point function (\ref{eq:DefD}) 
\bea
 D_{\sigma\lambda}(t,t')= F_{\sigma\lambda} (t,t') - {i \over 2 }  
 \rho_{\sigma\lambda}  (t,t') \textrm{sign}_C(t-t') ,
\eea
where 
\bea
 F_{\sigma\lambda} (t,t') &=& {1\over 2 } \langle  [ d_\sigma(t) , d^\dagger_\lambda (t') ] \rangle , \\ \nonumber
\rho_{\sigma\lambda} (t,t') &=& i \langle \{ d_\sigma (t) , d^\dagger _\lambda (t') \} \rangle ,
\eea
are the so-called statistical and spectral correlation functions in real-time representation. 
$ \textrm{sign}_C(t-t') = -1 + 2 \Theta_C(t-t') $ is a sign function on the 
contour.
For the equal-time arguments the 
following identities hold
\bea F_{\sigma\sigma}(t,t)= {1\over 2} - n_\sigma(t) , \ \ \ \ \ \ 
   \rho_{\sigma\lambda} (t,t) = i \delta_{\sigma\lambda} 
\eea
where $n_{\sigma}(t)$ is the mean number of fermions occupying the dot level with spin $\sigma$ at time $t$ and the equation for $\rho(t,t)$ follows from the fermionic equal-time anticommutation relations.
The symmetry relations
\bea F_{\sigma\lambda}(t,t') = F_{\lambda\sigma}(t',t) ^ *  , \ \ \ \ \quad 
\rho_{\lambda\sigma}(t,t')= - \rho_{\sigma\lambda}(t',t) ^ *.
\eea 
are a direct consequence of the definitions of $F$ and $\rho$.
For its numerical solution it is convenient to rewrite the Dyson equation for $ D(t,t')$ in terms of two equations for 
$ F(t,t')$ and $\rho(t,t')$ whereby the time integrals over the contour $\cal C$ are replaced by simple integrations along the real-time axis. 
In this way one arrives at the usual 2PI or Kadanoff-Baym 
equations of motion for the complex functions $F(t,t')$ and $\rho(t,t')$.

 We will assume throughout that 
correlations between up and down spins vanish, such that 
$  D_{\sigma\lambda}  \sim \delta_{\sigma \lambda} $.
We also introduce the notation: $D_{\sigma\sigma} \equiv D_\sigma  $  (no summation!). With this, the equations of motion read
\bea \label{eom_diag}
(i\partial_t - M_{\sigma}(t)) ~ \rho_\sigma(t,t') 
&= &   \int_{t'}^t du ~ \Sigma_\sigma^\rho
(t,u) \rho_\sigma (u,t') ,\nonumber\\
(i\partial_t - M_{\sigma} (t) ) F_\sigma (t,t') 
& =&  \int_0^t du 
~\Sigma^\rho_\sigma (t,u)
 F_\rho(u,t')  \nonumber \\
 & & - \int_0^{t'} du \Sigma^F_\sigma (t,u) \rho_\sigma(u,t'),
\eea
where the decomposition 
\bea \Sigma_\sigma(t,t') &=& -i \Sigma_\sigma^{(0)} (t) \delta (t-t') + \\ \nonumber
&& +
 \Sigma^F_\sigma(t,t') - {i \over 2 } 
\textrm{sign}_C (t-t') \Sigma^\rho_\sigma (t,t'). 
\eea
of the self-energy into local, $F$ and $\rho$ type terms has been used.
The time-local energy term in (\ref{eom_diag}) is given by
\bea M_\sigma(t)= E_{0\sigma} + \Sigma_\sigma^{(0)}(t)
\eea
and includes the mean-field shift originating from the double-bubble diagram in Fig~\ref{fig:loops}a.
The above integro-differential equations are equivalent to 
the exact Kadanoff-Baym equations and include higher-order
correlations through the non-Markovian memory integrals on
their right hand side. 

Including the second-order `basket-ball« diagram in Fig~\ref{fig:loops}a beyond the mean-field approximation 
yields a perturbative small-coupling approximation of $\Gamma_2[D]$,
\bea 
&&\Gamma_2 [D] = - U \int_C dt ~ D_\uparrow(t,t) D_\downarrow (t,t) 
 \nonumber\\
  && \quad+\ 
  {i \over 2 } U^2 \int_C dt\, dt' D_{\uparrow } (t,t') 
 D_{\uparrow } (t',t)  D_{\downarrow} (t,t')  D_{\downarrow} (t',t).\qquad
\eea
In the self-energy this corresponds to taking into account the tadpole and sunset 
diagrams.

\subsection{Mean-field approximation}
\label{meansec}

The one-loop contribution to the self-energy
gives the mean-field approximation of the dynamic equations (\ref{eom_diag}),
\bea 
\label{eq:SigmaMF}
\Sigma^{MF} _\sigma (t,t') =  i U \delta (t-t') D_{-\sigma} (t,t).
\eea
Note that $D_\sigma(t,t')$ has a jump at equal
time arguments. 
By analyzing the path-integral construction one finds that using the specific 
operator ordering in the Hamiltonian 
(\ref{eq:andersonh}) implies that the self-energy takes the form
\bea \label{meanfieldsigma}
\Sigma^\textrm{MF}_{\sigma} (t,t') &=& - i U n_{-\sigma} \delta (t-t').
\eea
See Appendix \ref{app:eqtp} for details.
The resulting equations of motion to this order of approximation are local in time and 
therefore Markovian. They constitute the 1-loop or mean-field 
approximation.

\subsection{Second-order approximation}
\label{basketsec}

We finally derive the order-$U^2$ contribution to the self energies, also called second-order Born approximation,
\bea \Sigma^\textrm{Sunset}_{\sigma} (t,t') = U^2 D_\sigma(t,t') D_{-\sigma} (t,t') D_{-\sigma} (t',t),
\eea
or, in components,
\bea  \label{originalsimgaeq} 
 \Sigma^{F}_{\sigma} &= &U^2 \left[ F_\sigma \left ( | F_{-\sigma} | ^2  - {1\over 4 } 
| \rho_{-\sigma} | ^2 \right)   \right. \nonumber\\ & &  \left. \qquad
 -{1\over 4 } \rho_\sigma ( F_{-\sigma} \rho_{-\sigma}^*  
+ F_{-\sigma}^* \rho_{-\sigma} )  \right] , \nonumber\\
\Sigma^{\rho}_{\sigma} &=& U^2 \left[  \rho_\sigma  \left ( | F_{-\sigma} | ^2  - {1\over 4 } 
| \rho_{-\sigma} | ^2 \right) + 
 \right. \nonumber\\ & &  \left. \qquad
+ F_\sigma  ( F_{-\sigma} \rho ^*_{-\sigma} +
     F_{-\sigma}^* \rho _ {-\sigma} ) \right] 
\eea
where we have omitted arguments.
The contributions (\ref{meanfieldsigma}) and (\ref{originalsimgaeq}) are summed up 
to be used in Eq. (\ref{eom_diag}), with  (\ref{meanfieldsigma}) contributing
 to $\Sigma^{(0)}_\sigma$. 
Some details concerning the numerical implementation of the above 
equations are given in  App. \ref{app:NumImp}.

\subsection{s-channel resummation}
\label{resumsec}

In this section we will summarize the $s$-channel resummation which leads substantially 
beyond the coupling approximations discussed before. 
Specifically, this involves a summation of the bubble chain contributions
with alternating spins where,
in each bubble, the two propagators describe the same spin component. 
This is similar to the next-to-leading-order $1/N$ approximation
for $N$-component scalar fields \cite{Berges:2001fi,Aarts:2002dj}. 
An elegant way to perform the resummation involves a Hubbard-Stratonovich (HS) transformation.
Since the interaction vertex couples 'up' spins with 
'down' spins, the bubbles in the chain have alternating spins. 
The action
\bea S_\mathrm{dot}= \int_C dt \sum_\sigma d^\dagger_\sigma ( i \partial_t -E_{0\sigma}) d_\sigma -
 U d^\dagger_{\uparrow} d_{\uparrow} d^\dagger_{\downarrow} d_{\downarrow}  
\eea
is rewritten using auxiliary scalar fields $\chi_1$ and $\chi_2$ by use of the substitution
\bea - J A^{-1} J \ \ \rightarrow \chi_T A \chi + 2 J_T \chi 
\eea
where 
\begin{displaymath}
J = {1 \over 2 }\left( 
\begin{array}{c}  d^\dagger_{\uparrow} d_{\uparrow}  \\ d^\dagger_{\downarrow} d_{\downarrow}  
\end{array} \right)
,\ \ \ \ 
\chi = \left( 
\begin{array}{c} \chi_1 \\ \chi_2 
\end{array} \right)
, \ \ \ \ A= {1 \over 2 U } \left(
\begin{array}{cc}
0 & 1 \\
1 & 0 \\
\end{array} \right).
\end{displaymath}
The resulting action reads
\bea \label{hsaction}
 S_\mathrm{dot} [ d_\sigma, d^\dagger _ \sigma, \chi_i ] &= & \int_C dt \sum_\sigma d^\dagger_\sigma 
( i \partial_t -E_{0\sigma}) d_\sigma  \nonumber\\ && 
+{1\over U } \chi_1 \chi_2 +  d^\dagger_{\uparrow} d_{\uparrow} \chi_1
  +  d^\dagger_{\downarrow} d_{\downarrow}    \chi_2 .
\eea
The free inverse propagators are read off from the quadratic part of the action
\bea i G_0^{-1}(t,t') &=& 2  A  \delta(t-t') , \nonumber\\
 i  D_{0,\sigma}^{-1} (t,t') &=&  ( i \partial_t - E_{0\sigma} + \bar \chi_i )
 \delta(t-t')
\eea
where the free propagator $G_0$  of the scalar fields is a $2 \times 2 $  matrix. 
Accordingly, we call  $G$  the propagator of the scalars, and 
$\bar \chi_i$ is the one-point function or expectation value of the auxiliary fields.

The corresponding 2PI effective action can be written as:
\bea
\Gamma [G,D,\bar \chi ] &=& S_{\text{dot}} [ d^\dagger_\sigma
= d_\sigma=0,\bar \chi ]   
   - i \textrm{Tr} \left[  \textrm{ln} D^{-1} + D_0^{-1} D 
\right] 
 \nonumber\\ && + {i\over 2 } \text{Tr} \left[ \textrm{ln} 
 G^{-1} + G_0^{-1} G \right]
 + \Gamma_2 [D,G] + \textrm{const.}\nonumber\\
\eea
where $\Gamma_2[D,G]$ contains all closed 
2PI diagrams built from the 3-point vertices 
of the action (\ref{hsaction}) and full scalar and fermion propagators.
The lowest-order contribution is shown in Fig.~\ref{fig:loops}b.

The stationarity conditions give the Schwinger-Dyson equations. In this case,
the field average of the scalar fields, in contrast to that of the 
fermionic fields, is nonzero, so we have stationary equations of the form
 $ \delta \Gamma / \delta \bar \chi = 0 $. The
resulting equations read 
\bea & \bar \chi_1 (t)  = U D_{\downarrow\downarrow} (t,t) ,
\qquad \bar \chi_2 (t)  = U D_{\uparrow \uparrow} (t,t) , &
\\ 
& D_0^{-1} D = \Sigma * D + \delta , & \\ 
& G_0^{-1} G = \Pi * G + \delta  & \label{constrainteq}
\eea
where we again suppressed the 
time arguments, and $*$ stands for convolution on the contour $\mathcal{C}$,
\bea
 (A*B) (t,t') = \int_\mathcal{C} dz ~ A(t,z) B(z,t').
\eea 
$\Sigma$ and $\Pi$ are the self-energy of the fermion and the 
boson fields, respectively:
\bea \label{resumsigma}
 \Sigma _ \sigma (t,t') &= &-i { \partial \Gamma_2[D,G] 
 \over \partial D_\sigma(t',t) }  = 
 -   D_\sigma(t,t') G_{\sigma\sigma}(t,t')  ,
\nonumber \\ \nonumber
 \Pi (t,t') &=&  2i { \partial \Gamma_2 [D,G]
 \over \partial G(t,t') } , \\ 
\Pi_{\sigma\sigma} &= & D_\sigma(t,t') D_\sigma(t',t) , \quad 
\Pi_{12} = \Pi_{21} = 0  
\eea
where $ \uparrow ~=1 , \ \downarrow ~=2 $ is used for the field indices.

The equations concerning the scalar fields are constraint equations, which 
do not contain any time derivatives,
because the $ \chi$ fields are auxiliary, non-dynamical fields.
The $\Gamma_2$ part of the action to lowest order is 
obtained as (See Fig. \ref{fig:loops}b)
\bea \Gamma_2 [D,G] = -{i\over 2} \sum_\sigma \int D_\sigma(x,y) D_\sigma(y,x) 
 G_{\sigma\sigma} (x,y).
\eea
From the constraint equation (\ref{constrainteq}) for $G$, one can see that
\bea
\label{eq:scalarProp}
  G &= & i U \delta \left( \begin{array}{cc}
0 & 1 \\
1 & 0 \\
\end{array} \right) 
- U^2 \left( \begin{array}{cc}
\Pi_{22} & 0 \\
 0  & \Pi_{11} \\
\end{array} \right)   \\ \nonumber  &&
- ~ i  U^3 \left( \begin{array}{cc}
0  & \Pi_{22} * \Pi_{11} \\
\Pi_{11} * \Pi_{22} & 0 \\
\end{array} \right)  \\ \nonumber &&
+ ~ U^4 \left( \begin{array}{cc}
\Pi_{22} * \Pi_{11} * \Pi_{22} & 0 \\
0 & \Pi_{11} *\Pi_{22} *\Pi_{11} \\
\end{array} \right) + \ldots ,
\eea
where we have again omitted the $(t,t')$ arguments.
Inserting $G_{11}$ and $G_{22}$ into the self-energy $\Sigma$ of the fermions
in eq. (\ref{resumsigma}), one finds a sum of bubble chains with alternating
spins being generated as mentioned initially.
The decomposition of the propagator $G$ and the self energies into statistical and spectral parts is deferred to Appendix \ref{app:DecompG}.

\section{Leads in the 2PI formalism}
\label{leadsect}

In this section we couple the single dot system to the leads taking
into account the terms (\ref{S:leads}) and (\ref{S:tunnel}) in the
action.  We can calculate the leads' contribution exactly, since
(\ref{S:leads}) and (\ref{S:tunnel}) only include terms up to
quadratic order, by integrating these out. 
The initial density matrix of the full system is assumed to be of the 
product form
 \bea \label{initialrho}
\rho(t=0) = \rho_\textrm{dot} \otimes\rho_\textrm{leads},
\eea
i.e., no correlations between dot and lead electrons exist initially. 
$\rho_\textrm{dot}$
is described in Sect. \ref{exacttimesect}, while $\rho_\textrm{leads}$ 
describes 
a grand-canonical ensemble, as shown below.
To include a thermal density matrix for the lead electrons, we  expand
their time-contour with a vertical part after $\mathcal C$ from $t_0$ to $ t_0
- i \beta $ for each momentum mode of the lead
electrons.  Then, the contribution to the dot degrees of freedom is
\bea 
i S_\textrm{env}= -i |\tau|^2 \int_C dt~ dt' d^\dagger (t) S^{-1}
(t,t') ~d(t') \eea 
where $S(t,t')$ is the free action of the lead
electrons, 
\bea S(t,t') = (i \partial_t - \epsilon (t)) \delta (t-t').
\eea 
Note that we have chosen symmetric tunneling $ \tau = t_L =t_R $.
Here $t$ is a complex variable, which lives on the extended
Schwinger-Keldysh contour. $\epsilon(t) = \epsilon_{kp} $ if $t$ is
real, with $k$ and $p$ being the indices of the mode considered, and $
\epsilon(t)= \epsilon_{kp}-\mu $ if $t$ is on the vertical part of the
contour.  This prescription gives the correct initial density matrix
corresponding to the grand canonical ensemble.

To obtain the contribution to $S_\textrm{env}$, we have to solve the 
equation
\bea ( i \partial_ t - \epsilon (t) ) A(t,t') = \delta_C (t,t'),
\eea
using the antiperiodic boundary condition $ A(0,t')=-A(-i\beta,t') $.
The result can be written in the form
\bea  \label{defa}
S^{-1}(t,t') &=& A(t,t') \nonumber\\ 
 &=& A^> (t,t') \Theta_C (t-t') - A^<(t,t') \Theta_C (t'-t ),\qquad
\\
A^>(t,t') &=& -i (1-f(\epsilon_{kp}-\mu)) e^{-i \epsilon_{kp} (t-t') },  \nonumber\\
A^<(t,t') &=& -i f(\epsilon_{kp}-\mu)  e^{-i \epsilon_{kp} (t-t') }
\eea
where $f(x) = 1/ [1 + \exp(\beta x )] $ is the Fermi function.
After integrating the lead electrons out, this appears as part of the free 
propagator of the dot electrons, so we have 
\bea 
 \left[  D_0^{-1}  +i A  \right]  * D = \Sigma * D + \delta.  
\eea
After decomposition one finds that the effect of one lead-electron mode
 can be seen as a contribution to the self-energy of the 
dot electrons $ \Sigma^F $ and $ \Sigma^\rho$:
\bea \label{oneleadsigmaf}
\Sigma^{F(1)}_\textrm{lead} (t,t')& =& - |\tau |^2 \left( {1 \over 2} 
 -  f( \epsilon-\mu)\right) e^{-i \epsilon (t-t')} , \\
\label{oneleadsigmarho}
\Sigma^{\rho(1)}_\textrm{lead}(t,t')&=& - i |\tau |^2 e^{-i \epsilon (t-t')}.
\eea
One can also argue the following way: If
the tunneling terms are used as a vertex, they give a one-loop 
diagram in  $\Gamma_2[D]$ , half of the loop is the propagator of the 
electron on the dot, the other half is the propagator of the lead-electron.
After taking the derivative of $\Gamma[D]$ one sees that
up to a factor the contribution to the
self-energy is the propagator 
of the free lead electron. This gives the same result as above.

Hence, the contribution only depends on the 
difference of the time arguments. This is to be expected, 
since the lead electrons 
are in thermal equilibrium.
Now we have the contribution of one lead particle. 
The wide-band limit corresponds to 
integrating over a continuum
\bea \Sigma_\textrm{lead}=\int_{-D}^D d \epsilon ~\rho (\epsilon) 
~\Sigma_\textrm{lead}^{(1)}  
\eea
with $\rho(\epsilon)= $ const. With (\ref{oneleadsigmarho}) this leads to
\bea 
 \Sigma^\rho_\textrm{lead} = - \int_{-D}^D d\epsilon ~ i | \tau |^2 e^{- i \epsilon t } = 
- i | \tau |^2  2 { \sin Dt \over t }.
\eea
It is useful to consider the limit $ D\rightarrow \infty $, where
\bea \Sigma^\rho_\textrm{lead} (t,t') = - 2 i \pi | \tau |^2 \delta( t-t') .
\eea

The calculation of $\Sigma^F$ can only be done analytically at zero 
temperature.
The finite-temperature correction 
is calculated numerically, whereas the zero-temperature contribution 
is a principal value,

\bea \Sigma^F_{\textrm{lead}, T=0} (t) = i |\tau |^2 P  { e^{-i \mu t} \over t } =
\sum_{L,R} \lim_{\epsilon \rightarrow 0}  i |\tau |^2   e^{-i \mu t} 
{ t \over t^2 + \epsilon^2 } .
\eea

To calculate the time dependent non-equilibrium current, one adds a 
further source term to the
action:
\bea S_\eta = {i e \eta \over 2 }  \sum_{kp\sigma} p ( t_p c^\dagger_{kp\sigma} d_\sigma - t^*_p 
d ^\dagger_\sigma c_{kp\sigma} ) (t _ m) 
\eea
Then the current at some measurement time $t_m$ can be written as 
\bea I(t_m) = -i \left.{ \partial \over \partial \eta } \ln Z [ \eta ] \right| _ {\eta=0} .
\eea
After integrating out the leads, one gets the following contribution 
to the action:
\bea
i S_\textrm{lead} = -i \sum_p \int _C dt \int _C dt' d^\dagger (t) \bigg\{ A_p(t,t')
 \nonumber\\  +
 { i e \eta \over 2 } A_p(t,t') \left[ - \delta (t-t_m) + \delta (t'-t_m)
\right]  \bigg\} d(t') 
\eea
where we dealt with the first term in the previous 
subsection, see Eq.~(\ref{defa}). 
The second term 
will provide a contribution to the 
current which thus reads
\bea 
 I_L(t) &=& e \sum _\sigma \int_ 0^{t_m} dt 
\Big[  \Big. A^F_{L}(t_m,t) \rho_\sigma (t,t_m)  \nonumber\\ 
&&\quad -\ A^\rho_{L} (t_m,t) F_\sigma(t,t_m) 
  \Big. + ~  A^{F* }_{L}(t_m,t) \rho_\sigma^*(t,t_m) 
\nonumber\\  
&&\quad  -\ A^{\rho* }_{L} (t_m,t) F_\sigma^*(t,t_m)
\Big]   . 
\eea
For the full current one needs $I=(I_L-I_R)/2$. If the leads are symmetric 
(tunneling rate, temperature and level density are the same for 
the left and the right lead) the $A_\rho$ terms cancel,
\bea
 I_\text{symm}= e \textrm{Re}  \sum_\sigma \int_ 0^{t_m} dt 
\left[ A^F_{L}(t_m,t)- A^F_{R}(t_m,t) \right] \rho_\sigma (t,t_m) .\
\eea

\section{Comparison with exact results}
\label{exasect}

\subsection{Exact time evolution without leads}
\label{exacttimesect}
Without the leads, the time evolution 
of the dot is a simple quantum-mechanics problem
of solving the Schr\"odinger equation. 
To compare with results obtained with the above functional methods,
 we need to choose an initial density matrix for which
the 1-point, 3-point and higher connected $n$-point functions are zero:
 $ \textrm{Tr}  ( \rho d_\sigma) = 0 , \ \  
  \textrm{Tr}  ( \rho d_\sigma d^\dagger_{-\sigma}d_{-\sigma})  = 0  , \ \ 
\textrm{Tr}  ( \rho F_{\sigma\lambda}) = \delta_{\sigma\lambda} (1/2-n_\sigma) , \ \ \textrm{Tr}  
( \rho  d_\sigma d_{-\sigma}) = 0 
$, etc.
These requirements fully constrain the density matrix.

Then one can solve the time evolution analytically by diagonalizing 
$H_\textrm{dot}$ defined in Eq. (\ref{andersonh}).
The commutator part of the 2-point function 
 is given by
\bea &&F_{11} (0,t) = F^*_{11} (t,0) = {1\over 2 }
\left[ (1-n_1)(1-n_2) e ^ { i E_{01} t }  \right.  \nonumber\\
 &&\quad\left.  + ~ (1-n_1)n_2 e ^ { i (E_{01}+U) t } 
 - n_1(1-n_2) e ^ { i E_{01} t }
  -n_1 n_2 e ^ { i (E_{01}+U) t } \right].\nonumber\\
\eea
%
\begin{figure}[t]
\begin{center}
\includegraphics[width=0.47 \textwidth]{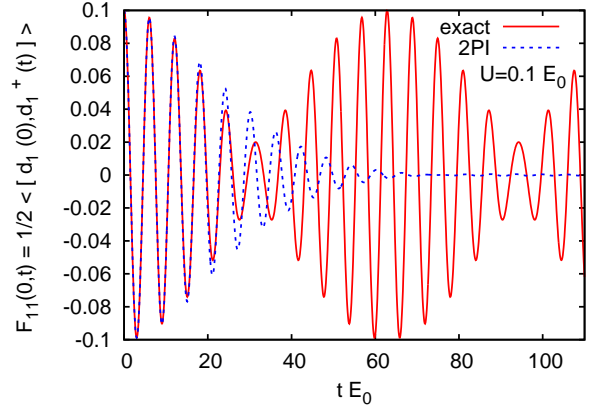}
\caption{ Time evolution 
of the two-time correlation function 
$F_{11}(0,t)$ for an isolated quantum dot: Comparison of 2PI and 
exact results obtained by directly solving the Schr\"odinger equation. The magnetic field is zero.
While the exact evolution shows revivals, the oscillations are damped out within the 2PI approach.
\label{fig:exact}}
\end{center}
\end{figure}
%
\begin{figure}[t]
\begin{center}
\includegraphics[width=0.47 \textwidth]{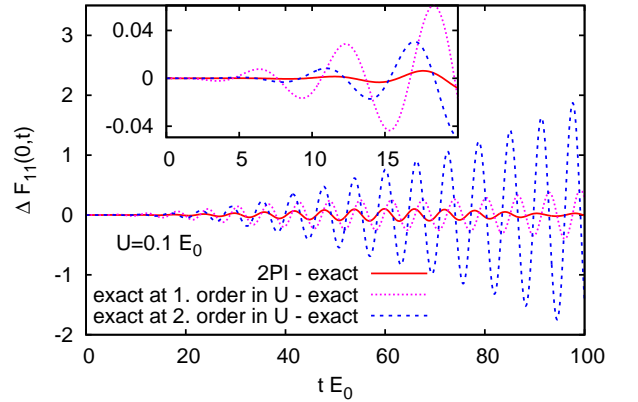}
\caption{ The same time evolution 
as in Fig.~\ref{fig:exact}: Comparison of 2PI and exact results. 
The difference between the exact and 2PI results, as well as the 
difference between the Taylor-expanded exact and the exact
 results are shown.
\label{fig:diff}}
\end{center}
\end{figure}
%
We compare this non-equal-time two-point function to the 
solution of the Kadanoff-Baym equations obtained from the 2PI effective action.
As one sees in Fig.~\ref{fig:exact}, the 2PI solution is in 
good agreement in the 
beginning, but does not show revivals as expected, cf., e.g., Refs.~\cite{Rey2004,Aarts:2006cv}.
Coupling to the leads causes a real damping which the 2PI approach is capable to describe. 
In Fig.~\ref{fig:diff} we plot the difference between the exact and the 
2PI results for $F_{11}(0,t)$, as 
well as the difference between the exact and the Taylor-expanded 
exact results. This illustrates that the 2PI equations yield a well-behaved dynamics.
In contrast to this, perturbation theory shows secularity problems.

\subsection{The current at $U=0$ }

Another quantity which can be calculated exactly is the current at vanishing on-dot coupling $U$.
This is due to the fact that at zero coupling the dot degrees of freedom are also quadratic. For
details of the calculation, see, e.g., Ref.~\cite{Gezzi2007}.
The result is
\bea \label{analcurrent} 
I= {1\over \pi} \int d\epsilon \left( f( \epsilon- \mu_L ) - f(\epsilon-\mu_R)
\right) { \Gamma ^2 \over ( \omega -E_0) ^2 + \Gamma^2 } 
\eea 
where $f(x) = 1 / [ 1+ \exp (\beta x) ] $ is the Fermi function.

\begin{figure}[t]
\begin{center}
\includegraphics[width=0.4 \textwidth]{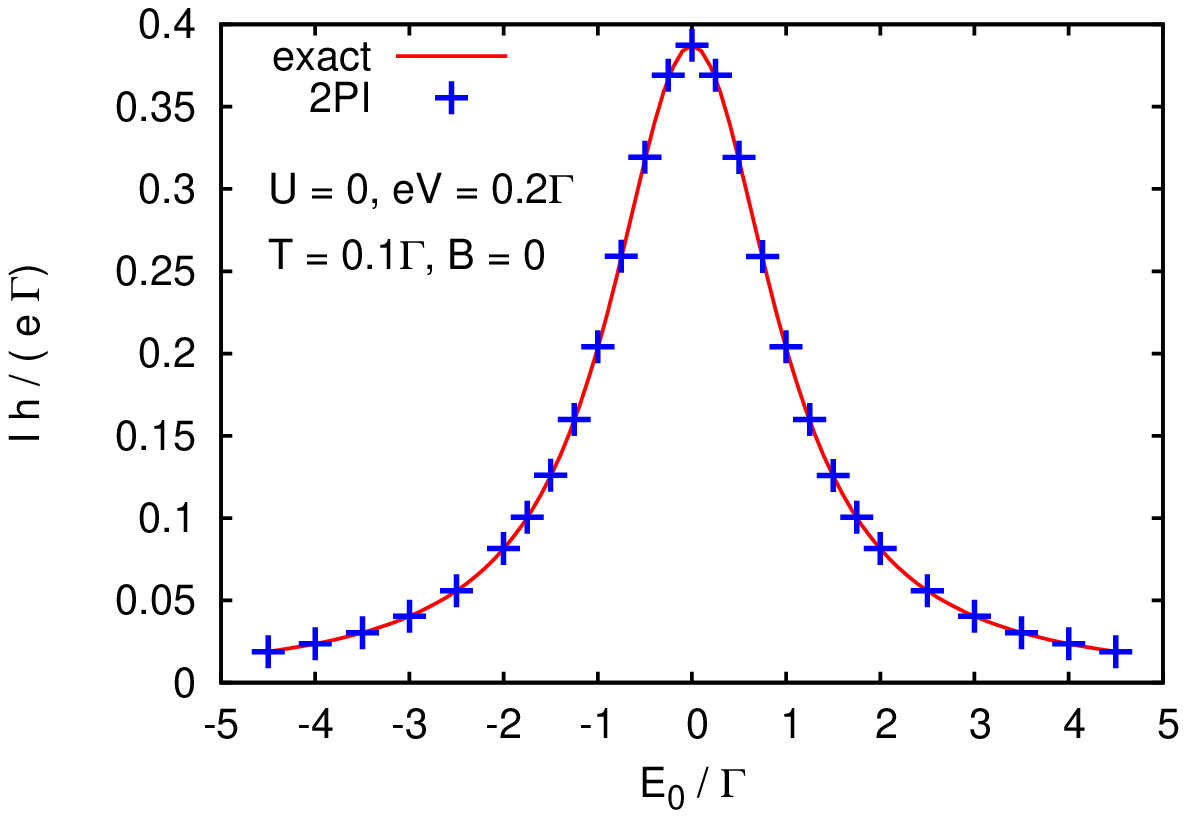}\\
\includegraphics[width=0.38 \textwidth]{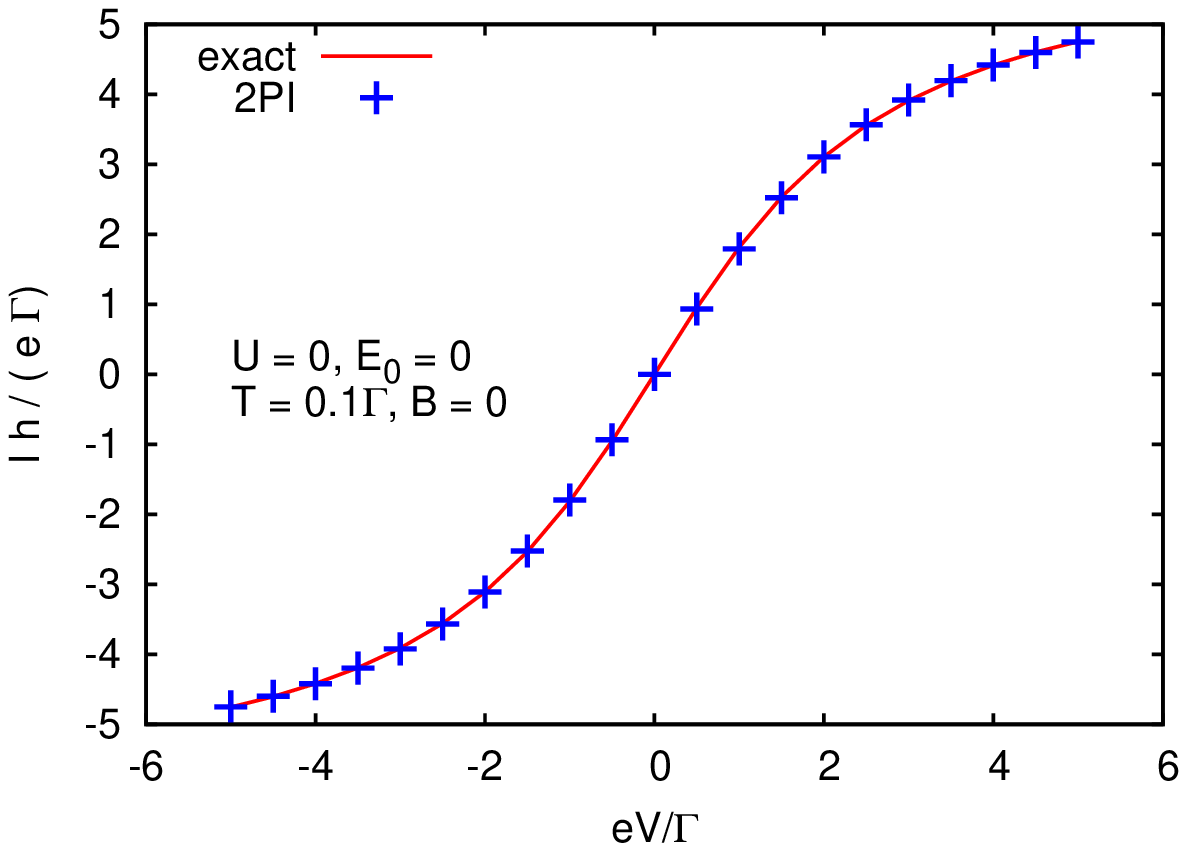}
\caption{The stationary current (\ref{analcurrent}) through the 
quantum dot at $U=0$, as a function of the bias voltage. 
The comparison of the 2PI results and the results of Ref.~\cite{Gezzi2007}, 
which in this case are both exact, serves as a benchmark test.
\label{fig:currentzerou}}
\end{center}
\end{figure}

In Fig.~\ref{fig:currentzerou}, the currents as obtained from the 
analytical formula (\ref{analcurrent})
and from the 2PI calculation (after waiting for the stationary state), 
for $U=0$, are compared. The agreement benchmarks our calculation, 
because we take the leads into account exactly, and at 
$U=0$ the 2PI treatment is also exact.

\section{Transient and stationary dynamics}
\label{ressect}

In the remainder of this paper we present our results for the transient build-up of the
current through the quantum dot, for different coupling strengths, bias voltages, and temperatures.
We compare our results with those obtained with alternative methods as given in the literature.
As we will show, these underline the necessity to go beyond the perturbative coupling expansion 
of the 2PI part $\Gamma_{2}$ of the effective action. 
Our results also give hints to the limitations of the $s$-channel resummation.
We consider first the transient evolution of the current, showing the build-up of the stationary 
state and then analyse the resulting asymptotic current.

\subsection{Numerical comparisons of different approaches}

For the comparisons we use data that was published in Refs.~\cite{ispi,eckel2010}. For details 
about the fRG (functional Renormalisation Group), ISPI (Iterative Summation of real-time 
Path Integrals), QMC (Quantum Monte Carlo) and
tDMRG (time-dependent Density Matrix Renormalization Group),
see, e.g., Ref.~\cite{eckel2010}, and references therein. 
If not otherwise stated, we will use the particle-hole symmetric point,
where $ E_0=-U/2 $, and $U=2 \Gamma$, $B=0$, $T=0.1\Gamma$.

\begin{figure}[t]
\begin{center}
\includegraphics[width=0.45 \textwidth]{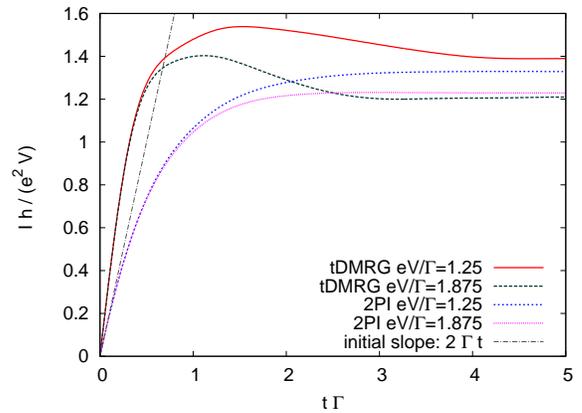}
\caption{ Comparison of the transient currents obtained 
with the 2PI equations and with the tDMRG method, at the 
symmetric point: $ E_0 = -U /2 $ .
The parameters are: $ U=3 \Gamma, T=0.1 \Gamma $. 
Note that the different initial slopes are due to differently chosen initial conditions.
While, in the tDMRG calculation, initial correlations between dot and leads are present, 
there are no such correlations in the 2PI approach.
\label{fig:transient}}
\end{center}
\end{figure}

In Fig.~\ref{fig:transient} we compare the time dependence of the
current through the dot as obtained with the full 2PI equations of motion with the
results of the tDMRG method. The stationary values obtained with the two methods
are close to each other, but the time dependences are different.  
The initial slope of the current for the 2PI curve 
is given by $ I h / (e^2  V  ) = 2 \Gamma t $,
as can be seen by evaluating the Heisenberg equation of motion $ \dot I = i / \hbar [ H, I] $,
with the uncorrelated density matrix (\ref{initialrho}) at the initial time. 
The tDMRG method uses an initial
density matrix containing correlations between dot and lead
electrons, which explains the different slope at the initial time.

\begin{figure}[t]
\begin{center}
\includegraphics[width=0.45 \textwidth]{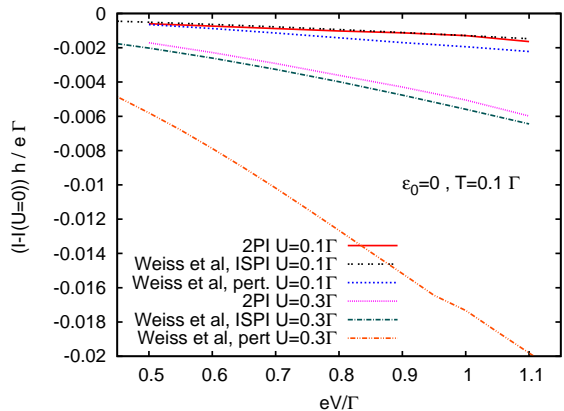}
\caption{The modification of the asymptotic 
stationary current due to nonvanishing interactions
is shown as functions of the bias voltage. The results obtained
with the 2PI method are compared with those of the ISPI calculations given in Ref.~\cite{ispi}, and with
perturbative results from Ref.~\cite{Dellanna2008}.
\label{fig:bugvolt}}
\end{center}
\end{figure}

In Fig.~\ref{fig:bugvolt} we compare, for small couplings, the stationary current with the corresponding results of
the ISPI approach discussed in Ref.~\cite{ispi}. 
The perturbative results were presented in Ref.~\cite{Dellanna2008}.
Note that we plot the interaction correction, that is, the
difference of the currents for zero and non-zero coupling which,
for the chosen couplings, is a small quantity.  The Coulomb-blockade
physics suggests that the sign of the correction should be negative, as
transport is suppressed for non-degenerate levels on the dot.
One observes that the
agreement between the results of the two methods, within the range of couplings considered, is very
good. For $U=0.1\Gamma$ the perturbative results are close to the ISPI
and 2PI results. We find that for $U=0.3\Gamma$ the perturbation theory
is of limited accuracy while the 2PI and ISPI results agree well.

\begin{figure}[t]
\begin{center}
\includegraphics[width=0.45 \textwidth]{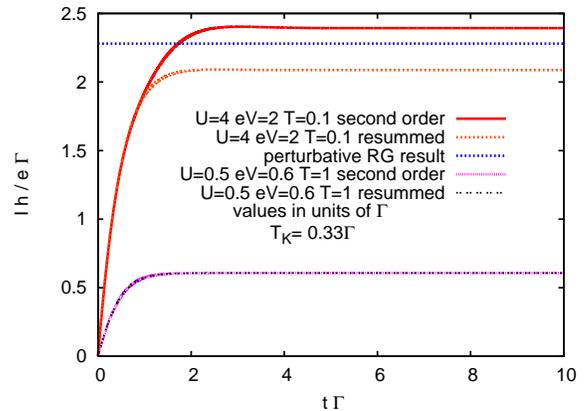}
\caption{The time dependent current through the dot is shown 
for different couplings $U$, bias voltage $V$, and temperature $T$. 
We compare results obtained within the second-order truncation of the 2PI effective action
with those derived after $s$-channel resummation. 
For $U=2 \Gamma$ there is perfect agreement. Note that, for the bigger coupling, the Kondo temperature 
(\ref{kondotemp}) is $T_K=0.33 \Gamma $ such that the system is already in the Kondo regime.
\label{fig:wp}}
\end{center}
\end{figure}

In Fig.~\ref{fig:wp} the time dependence of the current is shown for different choices of the interaction,
the bias voltage, and the temperature.
We compare the second-order coupling expansion, see Sect.~\ref{basketsec}, with the $s$-channel resummation introduced in Sect.~\ref{resumsec}.
For the case of stronger interactions we compare the stationary current reached with the result 
obtained with perturbative RG methods, as given in Ref.~\cite{ispi}. 
While for small coupling the truncations agree and reproduce the ISPI result
given in Ref.~\cite{ispi}, deviations appear for larger $U$,
where the system is expected to enter the Kondo regime.

\begin{figure}[t]
\begin{center}
\includegraphics[width=0.45 \textwidth]{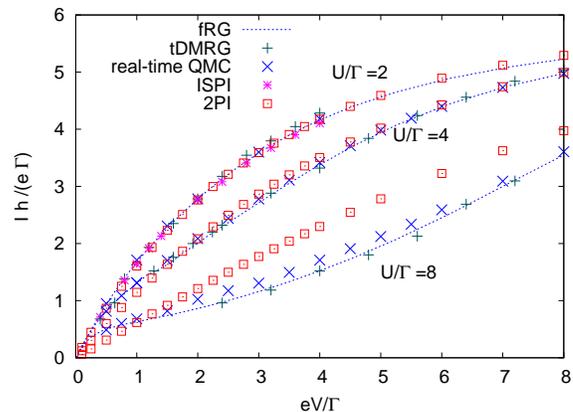}
\caption{ Comparison of the results obtained with different methods, at the symmetric point $E_0=-U/2$. 
The stationary current is shown as a function of the bias voltage for 
several coupling strengths. The 2PI and the ISPI methods 
use $T=0.1\Gamma$.
\label{fig:compf2}}
\end{center}
\end{figure}

In Fig.~\ref{fig:compf2} we compare the stationary current 
for bigger couplings, at the particle-hole symmetric point ($E_0 = -U/2$),
with corresponding results from the other methods.
With this we provide an add-on 
to the comparison presented in Ref.~\cite{eckel2010}. One finds that for $U=2\Gamma$
all methods agree. At $U=4\Gamma$ deviations start to appear 
which become larger at $U=8 \Gamma$.

\begin{figure}[t]
\begin{center}
\includegraphics[width=0.45 \textwidth ]{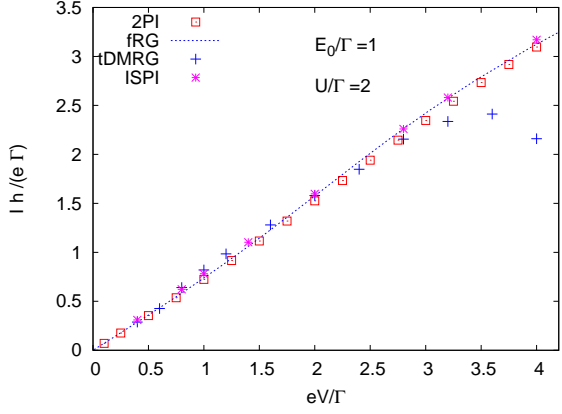}
\caption{ The stationary current in the mixed-valence regime for the coupling
of $U=2\Gamma$ , and $E_0 = \Gamma $. See Ref.~\cite{HM2009} of a detailed discussion of the
other methods.
\label{fig:mixed}}
\end{center}
\end{figure}

In Fig.~\ref{fig:mixed} we test the mixed-valence
regime, by comparing the steady-state current, using 
$E_0 = \Gamma= -U/2 + 2 \Gamma $. For small bias voltages one sees good 
agreement between all methods.
For further details concerning the other methods see the discussion in Ref.~\cite{HM2009}.
%
\begin{figure}[t]
\begin{center}
\includegraphics[width=0.45 \textwidth ]{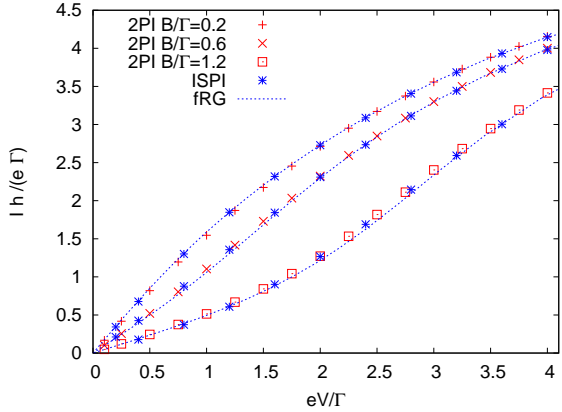}
\caption{ The stationary current  as a function of the 
bias voltage, for several magnetic field strengths.
\label{fig:bdep}}
\end{center}
\end{figure}
%
We furthermore compare our results with those from the fRG and ISPI methods at non-zero magnetic field, 
as shown in Fig.~\ref{fig:bdep}, and find good agreement for the full 
range of bias voltages.
%
\begin{figure}[h]
\begin{center}
\includegraphics[width=0.45 \textwidth ]{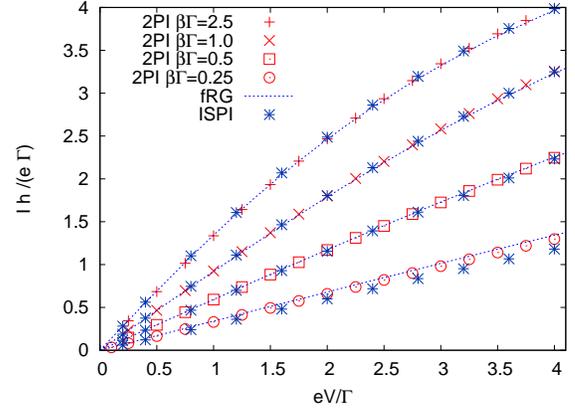}
\caption{  The stationary current is plotted as a function of the 
bias voltage for several temperatures.
\label{fig:tempdep}}
\end{center}
\end{figure}
%
In Fig.~\ref{fig:tempdep} we finally compare the temperature dependence 
of the steady-state current with fRG and ISPI results, for 
$U=2 \Gamma, \ \ T=0.1\Gamma$. One finds good agreement for all three methods, 
with the largest deviations in the high-temperature large-bias regime.

Overall, we conclude that in the moderate-coupling region, the 2PI method gives 
reliable results irrespective of the bias voltage and the temperature, both at
the symmetric point ($E_0=-U/2$) and in the mixed-valence regime.

\subsection{Effective coupling for $s$-channel resummation}
\label{sec:EffU}

We close this section with an analysis of the effect of the $s$-channel resummation on the 
effective coupling strength on the dot.
This resummation allows us to write the equation for 
the fermion self-energy as 
\bea \label{effsigma}
\Sigma^F &=&  U ( F A^F 
 - {1\over 4 } \rho  A^\rho ) =  U U_\textrm{eff} ( F \Pi^F - {1\over 4 } 
 \rho \Pi^\rho ) ,
 \nonumber\\
\Sigma^\rho &= & U ( \rho  A^F  + F A^\rho ) 
  =  U U _\textrm{eff} ( \rho  \Pi^F + F \Pi^\rho ) .
\eea
where 
\bea
\label{eq:EffU}
U_\textrm{eff}(\omega) = { 1- |\Pi^R |^2  \over | (\Pi^R)^2-1 |^2 }.
\eea
is a real, $\omega$-dependent effective coupling function.
Details of its derivation are given in Appendix \ref{app:EffU}. 

Note that, before resummation, $\Sigma^F$ and $\Sigma^\rho$ had the same 
form, with $U_\textrm{eff}$ replaced by $U$, 
see in Eqs. (\ref{originalsimgaeq}) and (\ref{srsigma}).
The time arguments in (\ref{effsigma}) are always $(t,t')$.

One can thus understand the effect of the resummation as introducing a 
frequency dependent 4-point vertex. This vertex only depends on one 
frequency, because only the $s$-channel diagrams
are resummed.
%
\begin{figure}[t]
\begin{center} 
\includegraphics[width=0.45 \textwidth]{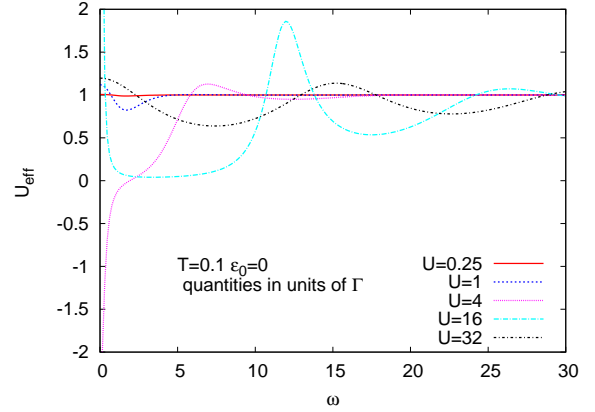}
\caption{The effective coupling for various couplings as a function 
of the frequency.
\label{fig:effcsat}}
\end{center}
\end{figure}
%
In Fig.~\ref{fig:effcsat} we show the effective coupling for various
bare couplings $U$.
For small couplings, one has $ U_\textrm{eff} \approx 1 $ because $\Pi$ is 
proportional to the coupling $U$. For medium-sized $U$, the 
effective coupling shows big deviations from $1$, and  
for large $U$ the effective coupling again approaches one.

\section{Conclusions}
\label{consect}

In this paper we have studied the real-time evolution of 
 the Anderson quantum dot, using the two-particle-irreducible (2PI) effective-action approach.
 The dynamical evolution of correlation functions is described by Kadanoff-Baym equations of motion 
 derived from the effective action in nonperturbative approximation.
Two different approximations were considered: the second-order (sunset) coupling-expansion 
approximation and the $s$-channel
 resummation approximation, in which the bubble chains (bubbles with equal-spin
propagators, alternating spins in adjacent bubbles) were summed to all orders.

The leads were taken into account by integrating them out exactly, using 
a grand-canonical initial density matrix. 
At nonzero bare interaction strength $U$ of the dot fermions we compared our results with those 
obtained within different other approaches. 
At moderate couplings, the 2PI method gives reliable results irrespective of the 
temperature or the bias voltage both in the particle-hole symmetric
and the mixed valence regimes.

Considering the effective coupling one finds that the 
resummation can be thought of as using a frequency dependent
four point vertex.
One way to go beyond the approximations of this study is to take into 
account a higher loop expansion of the 2PI part $\Gamma_2[D]$ of the 
Hubbard-Stratonovich transformed theory. While this requires more numerical 
effort, since the self energies in higher loop approximation include 
at least one inner vertex, for which an integration has to be carried out,
it is nevertheless possible with current computers.

An expansion of the presented scheme to 
the non-equilibrium RG approach put forward in Refs.~\cite{Gasenzer:2008zz,Gasenzer2008a} is under development.
It was shown that in the $s$-channel truncation this RG scheme is equivalent to 
the 2PI scheme used in this work. 
Beyond this one can take into account resummations also in the $t$ and $u$ channels,
also called spin-singlet or electron-hole and electron-electron channels, respectively, 
which are expected to yield the Kondo-resonance behaviour for large
$U$ \cite{Janis2007,Janis2008,Aug2010}.

\begin{acknowledgements}
The authors would like to thank J.~Berges, T.~Enss, F.~Heindrich-Meisner, 
S.~Jakobs, and H.~Sch\"oller for inspiring and useful discussions, 
J.~Eckel for providing the data published in Ref.~\cite{eckel2010}.
They acknowledge the support by the Deutsche Forschungsgemeinschaft, as well as the support by the Alliance Programme of the Helmholtz Association (HA216/EMMI), and by the Excellence Programme FRONTIER of the University of Heidelberg.
A large part of the numerical calculations for this project were done
on the bwGRiD (http://www.bw-grid.de), member of the German D-Grid
initiative, funded by BMBF
and MWFK 
Baden-W{\"u}rttemberg.
\end{acknowledgements}

\begin{appendix}

\section{Equal-time propagator}
\label{app:eqtp}

The one-loop contribution to the self-energy
gives the mean-field approximation (\ref{eq:SigmaMF}) of the dynamic equations (\ref{eom_diag}).
Note that $D_\sigma(t,t')$ has a jump at equal
time arguments. Hence and one has the choice between $D_{-\sigma}
(t,t+\epsilon)$ and $D_{-\sigma} (t,t-\epsilon)$, or some linear
combination of them.
\bea   \label{tadpoleord} \lim _ {\epsilon \rightarrow +0 } D_{\sigma} (t,t-\epsilon) = 
 d_\sigma(t) d_\sigma^\dagger(t)= 1 - n_{\sigma}(t),  \nonumber\\
  \lim _ {\epsilon \rightarrow +0 }  D_{\sigma} ( t, t+\epsilon) = 
-d^\dagger_\sigma (t) 
 d_\sigma (t) =  -n_{\sigma}(t) 
\eea
It is well known that 
this ambiguity in the 2 point correlation function corresponds to different 
orderings in the operator language.  For example, choosing the second
possibility in (\ref{tadpoleord}) corresponds to the operator Hamiltonian
\bea \label{eq:andersonh}
 H = \sum_\sigma E_{0\sigma} d^\dagger_\sigma d_\sigma   + 
U d_\uparrow ^\dagger d_{\uparrow} d_{\downarrow} ^\dagger d _\downarrow .
\eea
This can be shown by using the fact that a symmetrically ordered 
Hamiltonian corresponds to a symmetrised path integral prescription. 
Thus, 
\bea
 H_\textrm{symm}& =& \sum_\sigma E_{0\sigma} d^\dagger_\sigma d_\sigma   + 
 {U \over 4 } 
( d_\uparrow ^\dagger d_{\uparrow} d_{\downarrow} ^\dagger d _\downarrow -
  d_\uparrow  d_{\uparrow}^\dagger d_{\downarrow} ^\dagger d _\downarrow \nonumber \\ \nonumber && + ~
  d_\uparrow  d_{\uparrow}^\dagger d_{\downarrow}  d _\downarrow ^\dagger -
  d_\uparrow ^\dagger d_{\uparrow} d_{\downarrow}  d _\downarrow ^\dagger) 
 \nonumber \\  & = &\sum_\sigma \left(E_{0\sigma}-{U\over 2 } \right)
d^\dagger_\sigma d_\sigma   + 
U d_\uparrow ^\dagger d_{\uparrow} d_{\downarrow} ^\dagger d _\downarrow .
\eea
is related to 
\bea \nonumber \Sigma^{MF}_{\textrm{symm},\sigma} & =& {1\over 2 } \left[  
\lim _ {\epsilon \rightarrow +0 } D_{-\sigma} (t,t-\epsilon) +
\lim _ {\epsilon \rightarrow +0 } D_{-\sigma} (t,t+\epsilon) \right] 
 \\ &=&  {1\over 2} - n_{-\sigma} = F_{-\sigma} (t,t).
\eea
By analyzing the equations of motion, one finds that this is equivalent to using the 
original Hamiltonian (\ref{eq:andersonh}), and the following 
prescription
\bea 
\Sigma^\textrm{MF}_{\sigma} (t,t') &=& i U \delta (t-t') \lim _ {\epsilon \rightarrow +0 }   
D_{-\sigma} (t,t+\epsilon)    \nonumber\\ &=& - i U n_{-\sigma} \delta (t-t').
\eea
For higher-order contributions no such complication of ordering is present.

\section{Numerical Implementation}
\label{app:NumImp}

The naive (Euler) discretisation of the equation
(\ref{simpleeq}) breaks down quickly, because the absolute value of 
$F(t,t')$ is not conserved, even for zero right-hand side $R$.
\bea (i \partial_t - E_0)  F(t,t') = R \label{simpleeq}
\eea 

One gets a much better discretisation if one first inserts an ansatz which
describes the free behaviour, and then discretises the equation 
for the remaining time evolution.

For this one introduces the new functions $H(t,t')$ and $s(t,t')$
\bea
F(t,t') = e^{ - i E_0 (t-t') } H(t,t'),  \quad  
\rho(t,t') = e^{ - i E_0 (t-t') } s(t,t'). 
\eea

Without interactions, the solution is $H(t,t')=\textrm{const}$ and 
$s(t,t')=\textrm{const}$. 
Writing the discretised equations for $H(t,t')$, and similarly 
for $s(t,t')$: 
\bea H(t+\Delta t ,t') =  -i e^{i E_0 (t-t') }
  R   H(t,t') \Delta t + H(t,t') ,
\eea
and multiplying this equation with  $  e^{-i E_0 (t+\Delta t-t') } $
one gets an equation for $F(t,t')$:
\bea \label{neweom}
 F(t+\Delta t,t') = -i e^{-i E_0 \Delta t } 
 R   \Delta t + F(t,t')  
e^{-i E_0 \Delta t }. 
\eea

This equation defines a discretisation of (\ref{simpleeq}) that is 
different from the usual Euler discretisation, but has the same
continuum limit.
We see that using this algorithm, the free evolution
is reproduced without numerical errors. Hence for a moderate right-hand 
side one can use a big timestep, and still conserve the unitarity
of the equation to a good degree.

Solving the 2PI equations, we 
set $\rho(t,t)=i$ 
explicitly, as prescribed by the the anticommutation relations.
 Everything else is 
calculated using the discretisation (\ref{neweom}) of the EOM. 
On the right-hand side of the EOM, the memory integrals 
for a particular time $t$ require 
field values in the past, which have been computed 
before. 
Therefore, our  
scheme allows an explicit calculation of the next timestep from 
known quantities. We typically use $\Delta t=0.01-0.02$, which gives 
the continuum limit to a good accuracy.

For zero magnetic field, when $E_{0\uparrow} = E_{0\downarrow} $, one can 
use a symmetric initial condition
to save numerical resources. These symmetries apply in the second order
truncation as well as in the resummed truncation. We discuss 
the resummed equations below: starting from an initial condition,
 where the following equations are satisfied for $t=t'=0$, then 
they hold for any $ t , t' > 0 $.  

\bea \label{symm_eq}
\bar G_{21} (t,t')= \bar G_{12}(t,'t) , \quad \bar G_{11}(t,t')
= \bar G_{22}(t,t'), 
\\ \nonumber \Pi_{11}(t,t') = \Pi_{22} (t,t') ,\quad D_\uparrow (t,t')
= D_\downarrow (t,t').
\eea

The real constraint set by these equations is only  
$D_\uparrow (0,0)= D_\downarrow (0,0)$, the rest follows automatically 
via the constraint equations (\ref{constrainteq}) of the scalar
fields and the 
EOM for the fermionic fields.

\section{Decomposition of the propagator $G$}
\label{app:DecompG}
According to (\ref{constrainteq}), $ G (t,t')  $ is decomposed as
$  G (t,t')= \bar G (t,t') + i U \sigma_1  \delta (t-t') $, where $\sigma_1$
is the first Pauli matrix.
The equation for $\bar G$ reads
\bea \label{sk_eom}
 \left( \begin{array}{cc}
\bar G _ {2 1}  & \bar G_{22} \\
 \bar G_{11}   & \bar G _{12} \\
\end{array} \right) 
= i U \Pi  * \bar G - U^2 
 \left( \begin{array}{cc}
 \Pi _{12}  & \Pi_{11} \\
 \Pi_{22}   & \Pi _{21} \\
\end{array} \right).
\eea
The fermionic and bosonic self energies can be decomposed in the usual way 
into $F$ and $\rho$ 
parts 
\bea \label{srsigma}
\Sigma^F_{\sigma} &=& -1 ( F_\sigma \bar G _{F\sigma\sigma} 
 - {1\over 4 } \rho_\sigma \bar G_{\rho\sigma\sigma} ),
\nonumber\\
\Sigma^\rho_{\sigma}& =& -1 ( \rho_\sigma \bar G_{F\sigma\sigma} 
+ F_\sigma \bar G_{\rho\sigma\sigma} ),
\nonumber\\
\Pi^F_{\sigma} &=&  |F_\sigma|^2 - {1\over 4} | \rho_\sigma| ^2 ,
 \nonumber\\
\Pi^\rho_{\sigma} &=&  2 \textrm{Re} (F^*_\sigma \rho_\sigma ) ,
\eea
where, similar to the fermions, the scalar propagator is decomposed into
\bea
\bar G_{ij}(t,t') = \bar G^F_{ij}(t,t') - {i\over 2 } \text{sign}_C(t-t') 
 \bar G^\rho_{ij}(t,t').
\eea
In this setup, the ordering problem also appears, 
in the equations for the field average of the $ \chi$ fields. It can be dealt 
with similarly as in the previous subsection: symmetric Hamiltonian 
corresponds to a 
symmetrised path integral. From this one deduces the prescription for the 
non-symmetrised Anderson Hamiltonian (\ref{andersonh}). Eq. (\ref{eom_diag}) 
corresponds to the ordering of (\ref{andersonh}) when using 
\bea M_\sigma(t) = E_{0\sigma} + U n_{-\sigma} (t). 
\eea
The scalar constraint equation (\ref{sk_eom}) can be decomposed 
using the identity
\bea (i A * B) (t,t') &=& \int_0^{t} dz
A^\rho (t,z) B^F (z,t') 
\nonumber\\ &-&  \int_0^{t'} dz A^F(t,z) B^\rho(z,t') 
\nonumber\\ &-&   {i\over 2} \text{sign}_C(t-t') \int _{t'}^t A^\rho(t,z)
B^\rho(z,t'). \eea

\section{Effective coupling for $s$-channel resummation}
\label{app:EffU}

In this appendix we give the details of the calculation of the effective coupling given in Section \ref{sec:EffU}.
In the zero-magnetic-field case, where $E_{0\uparrow}=E_{0\downarrow}$, and
the symmetries of (\ref{symm_eq}) hold, one can define:
\bea A (t,t') & = & { \bar G_{11} (t,t') \over U }, 
\qquad B (t,t') = { \bar G_{21} (t,t') \over U }, \nonumber\\ 
\Pi (t,t') &= &U \Pi_{11} (t,t').
\eea
Hence, the constraint equations (\ref{constrainteq}) can be written as (suppressing time arguments)
\bea A =  i \Pi * B - \Pi  \label{eq:trinvresum} , \qquad 
  B = i \Pi * A .
\eea
After the transient processes are finished, in the stationary state, 
all two-point 
functions will only depend on the differences of the time coordinates.
In this case we can push the initial time to negative infinity, and 
 we can use 
the following decomposition identities for convolution on the 
Schwinger-Keldysh contour starting at $t_0=- \infty $:
\bea \label{decompid}
 i ( X * Y )^F & =&  X^R * Y^F - X^F * Y^A , \nonumber\\
     i ( X * Y ) ^\rho &=& X^R * Y^\rho - X^\rho * Y^A , \nonumber\\
     i ( X * Y )^R &=& X^R *Y^R ,\nonumber\\
     i ( X * Y )^A &=& - X^A *Y^A 
\eea
where the retarded and advanced two-point functions are defined as
\bea G^R(t,t') &=& \theta (t-t') G^\rho(t,t'),  \nonumber\\
G^A(t,t') &=& \theta (t'-t) G^\rho(t,t') .
\eea

Using this in Eq.~(\ref{eq:trinvresum}), after eliminating $B$ 
and Fourier transformation with respect to $t-t'$, we obtain
\bea 
 A^F&=& \Pi^R \Pi^R A^F  - \Pi^R \Pi^F A^A - \Pi^F 
+ \Pi^F \Pi^A A^A ,  \nonumber\\ 
 A^\rho &=& \Pi^R \Pi^R A^\rho  - \Pi^R \Pi^\rho A^A - \Pi^\rho 
+ \Pi^\rho \Pi^A A^A 
\eea
where we have omitted arguments ($\omega$). 
With the help of (\ref{decompid}), Eq. (\ref{eq:trinvresum}) can be rewritten
as
\bea 
\left( 1 + \Pi^R A^A - \Pi^A A^A \right) \left( \Pi^A \Pi^A -1 \right) 
= \Pi^R \Pi^A -1 .
\eea
We can write 
\bea A^\rho = - \Pi^\rho U_\textrm{eff}. \qquad 
A^F = - \Pi^F U_\textrm{eff} 
\eea
where
\bea
U_\textrm{eff}(\omega) = { 1- \Pi^R \Pi^A  \over ( (\Pi^R)^2-1 ) ( (\Pi^A )^2 -1 ) }
= { 1- |\Pi^R |^2  \over | (\Pi^R)^2-1 |^2 }.
\eea
which is the expression given in Eq.~(\ref{eq:EffU}).

\end{appendix}



\end{document}